\documentclass[preprint,10pt,plainnat]{aastex6} 
\usepackage{amsmath,amstext} 
\usepackage[figure,figure*]{hypcap} 
\usepackage{newtxmath} 
\usepackage{diagbox}

\begin{document}

\title{ Boundary conditions for similarity Index}
\author{Madhu Kashyap Jagadeesh \altaffilmark{1} and Pursharth Saxena\altaffilmark{2}}
\email{kas7890.astro@gmail.com}
\altaffiltext{1}{Department of Physics, Jyoti Nivas College, Bengaluru-560095, Karnataka, India}
\email{saxena.puru005@gmail.com}
\altaffiltext{2}{Department of Mathematics, Christ (Deemed to be University), Bengaluru 560 029, India}

\begin{abstract}
The recent development, shows that the Bray-Curtis's formula for similarity Index (1957), has been applied in various fields like Ecology, Astrophysics, etc. In this paper, we found the possible boundary conditions for this evolved formula (\textit{i.e} the numerical range in which the formula becomes in-effective to give the expected result). Here we have simulated the real world data in the form of normally distributed random numbers, that directly shows the range (or conditions) at which this formula gives unambiguous similarity result. 

\end{abstract}

\keywords{Similarity Index, Metric Tool, Boundary Conditions}


\section{Introduction}
Similarity Indexes are used to compare two or more quantitative/numeric data set, such as: the variation from a reference state. These indices are applied for classification of objects, clustering and retrieval problems, or to compute the overlaps between quantitative data \citep{Madhu}. The range of the index is defined on the closed interval of [0,1], where 0 and 1 corresponds to absolute dissimilarity and absolute similarity respectively. According to \citep{Bos}, from ecological studies, there are many multivariate techniques like similarity index, but not many of them are preferred. Since this one is widely accepted and used, we have chosen a random data set which disapproves with the above said agreement, with certain boundary conditions. There are many techniques applied in the ecological community (Such as: \citep{Aus}, \citep{or}, and \citep{And}), and according to \citep{BrayCurtis} there will be distortion in the data set. Later in 2011 the similarity index reached extra-terrestrial research, like finding Earth-like planets using Earth Similarity Index (ESI) \citep{schu} and for Mars-like planets applying Mars Similarity Index (MSI) \citep{Madhu}.\\

The structure of the paper is as follows: section 2 contains Mathematical formulation, and section 3 follows up with assumptions and graphical analysis, the last section is discussion and conclusion.\\

\section{Mathematical formulation of Bray-Curtis Similarity Indices}

Similarity between samples having multivariate data was well established by Deza and Deza in 2006. The practical application of the same (with ecology community) was done by Looman and Campbell, 1960.
This substantiates the usage of Bray--Curtis' work of 1957, which can be mathematically expressed as:

\begin{equation}
d_{\rm BC} = {\sum_{i=1}^{n}|p_i-q_i|  \over \sum_{i=1}^{n}\left(p_i+q_i\right)}\,.
\label{SIIeq2}
\end{equation}
Where $p_i$ and $q_i$ are measurable quantities between which the distance is to be measured, and $n$ is the total number of variables (Kashyap et al. 2017).
 Later in 1981 Bloom's transformation technique for the similarity measures for intersection was described as:

\begin{equation}
s_{\rm BC} = 1- d_{\rm BC} =
1- {\sum_{i=1}^{n}|p_i-q_i|  \over \sum_{i=1}^{n}\left(p_i+q_i\right)}\,.
\label{eq:similarity}
\end{equation}
Here, $0$ conveys the complete absence of relationships, whereas, $1$ shows a complete match of the two data records in the $n$-dimensional space \citep{JanSchulz}.

For heterogeneous data, a process of standardization, \textit{i.e.} balancing and equating the contribution of different types of variable, is needed to find the Similarity (Greeenacre and Primicerio 2013).

Traditionally, the similarity indices are subdivided into equal 0.2 intervals \citep{Bloom}, defining very low, low, moderate, high and very high similarity regions. Therefore, the threshold can be defined as:
\begin{equation}
V= \left[1-\Big|\frac{x_{ref}-x}{x_{ref}+x}\Big| \right]^{w_x}\,.
\label{SIIeq6}
\end{equation}
where $x$ can be any physical entity (such as: Mass, Concentration, Radius, etc...), $w_x$ is the weight for that particular entity, $x_ref$ is the reference value, and the dimension $n=1$. Usually for formulating similarity index, threshold $V=0.8$.

We define the physical limit as $x_a$ and $x_b$, which defines the limit for variation of the variable with respect to $x_{ref}$ (i.e. $x_a<x_{ref}<x_b$). Weight exponent are then calculated for the lower limit, $w_a$ and the upper limit, $w_b$
\begin{equation}
w_a = \frac
{\ln{V}}
{\ln\left[1-\left|\frac{x_{ref}-x_a}
{x_{ref}+x_a}\right|\right]}
\,,\quad 
w_b = \frac
{\ln{V}}
{\ln\left[1-\left|\frac{x_b-x_{ref}}{x_b+x_{ref}}\right|\right]}
\,,\quad 
\label{eq:weight_exponent}
\end{equation}
The average weight is found by the geometric mean,
 \begin{equation}
w_x=\sqrt{{w_a}\times {w_b}}\,.
\label{eq:geom_mean}
\end{equation}

In this paper, we use Eq.~\ref{eq:similarity} to define the similarity index 
\begin{equation}
SI_x = {\left[1-\Big|
\frac{x-x_{ref}}{x+x_{ref}}\Big| \right]^{w_x}}\,,
\label{eq:esi}
\end{equation}
The assumption in the Bray-Curtis scale is that samples are taken from same physical  measure, e.g. mass, or volume. And the final form of similarity index obtained satisfies the Bray-Curtis assumption.\\

\section{Assumptions for boundary conditions}
Let $R = {\left\lbrace{a_0, a_1, a_2 \dots a_m}\right\rbrace} \quad m \in \mathbb{N}$ be a non-empty set containing raw data.\\
Then, Consider a non-empty set, $S = {\left\lbrace{x_0, x_1, x_2 \dots x_n}\right\rbrace} \quad n \in \mathbb{N}$ to be a subset of R, such that $n \leq m$.\\

The weight exponents for calculating Similarity Index, and accounting for the Similarity Index itself, is dependent on $x_{min}$, $x_{max}$, and $x_{ref}$, \textit{i.e} the minimum, maximum, and reference value respectively. Where $x_{min}, \quad x_{max} \in R$ and $x_{ref}$, is the pre-defined reference point, basically set up by the application requirement, this can also belong to the set $S$ or set $R$.\\

Basic range definitions of the above mentioned parameters are:
\begin{align*}
x_{min} \in [0,9]\\
x_{max} \in [100,999]\\
x_{ref} \in [10,99]
\end{align*}

Then we define the boundary condition as
\begin{equation}
x_i \leq x_{ref} \forall x_i \in S
\label{eq:ineq_met}
\end{equation}
Similarly, 
\begin{equation}
x_i > x_{ref} \forall x_i \in S
\label{eq:ineq_not_met}
\end{equation}

In the above expressions, Eq (\ref{eq:ineq_met}) gives \textit{expected value} for the similarity index analysis. However, Eq (\ref{eq:ineq_not_met}), on the other hand, \textit{does not yield} the expected value. This pattern was observed in the raw data in the form of concentration of a trace gas (volume). Which satisfies the assumption established by Bray \& Curtis, 1957.

\end{}

\newpage 

\section{Results}
The patterns involving the inequality in Eq (\ref{eq:ineq_not_met}) were observed while dealing with concentration of atmospheric gases. Since the application part of the paper has not been published, that data-set is deemed to confidential, and hence random data samples were generated based on the same pattern as it was seen in the concentration of that specific trace gas. The reference for calculating the weight exponents using python source code is made available in appendix section.
\\

Here $2000$ random samples (set $R$) were drawn form a Normal Distribution having mean = $2$ and variance = $1$. The samples were then divided into five parts and average value of these five parts was taken as the set $S$. All the samples were multiplied by a weight of 100 for the different case studies, the corresponding graphical analysis is shown below. 

Since the reference value $x_{ref}$, is an inherent property of the parameter under study, and not the equation, $220$ and $50$  were randomly chosen to meet the conditions of Eq (\ref{eq:ineq_met}) and Eq (\ref{eq:ineq_not_met}) respectively. In Table 1, $x_{ref}$ is the reference value for which the conditions satisfying and dis-satisfying similarity index are observed. $X = S$ is the set as described above.

\begin{center}
	\centering
    \textbf{Table 1: Similarity index projections}
    \label{tab:primary}
	\begin{tabular}{ |p{3cm}|c|c|c|c|c| } \hline
		\textbf{Array Index $\rightarrow$} & \textbf{0} & \textbf{1} & \textbf{2} & \textbf{3} & \textbf{4}\\\hline
		\textbf{$X$} & 200.35 & 200.51 & 194.24 & 206.88 & 207.21 \\\hline
		\textbf{Similarity Index ($x_{ref} = 50$)} & 0.91687 &  0.91681 &  0.91901 &  0.91464 & 0.91453 \\\hline
        \textbf{Similarity Index ($x_{ref} = 220$)} & 0.99327 &  0.99332 &  0.99098 &  0.99560 & 0.99572 \\\hline
	\end{tabular}
\end{center}
Here the 2nd row of the table represents the condition where similarity index metric tool does not yield the expected value. Hence we define the boundary conditions as suggested in the assumption.\\

{\it\underline{{Graphical Analysis:}}}\\

\begin{figure}[h!]
\centering        
\includegraphics[width=10cm,angle=0]{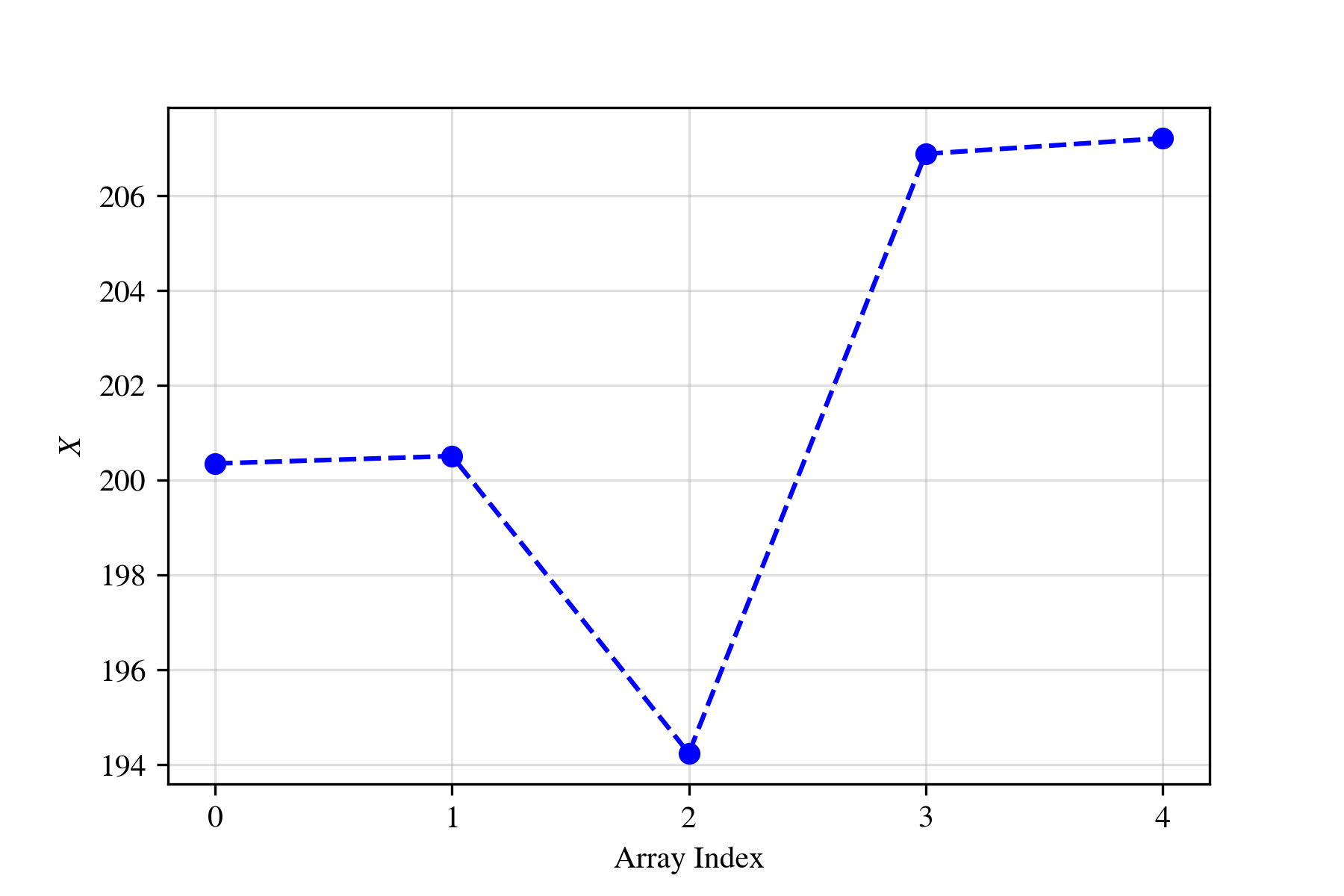} 
\caption{Values $x_i \in X$ with respect to $x_i$'s position in the array}
\label{fig:0}
\end{figure}

 In Fig \ref{fig:0} the values of $X$ are plotted. The abscissa in this particular case study is taken as the position of $x_i \in X$ in the array. \textit{i.e} for abscissa 0 the corresponding value of ordinate is: $x_{0}$, and so on. In practice abscissa for the similarity index can be anything, eg. mass, concentration, etc, which again is an inherent property of the parameter under observation and therefore, taking abscissa as array index does not interfere with the results obtained.

\begin{figure}[h!]
\centering        
\includegraphics[width=15cm,angle=0]{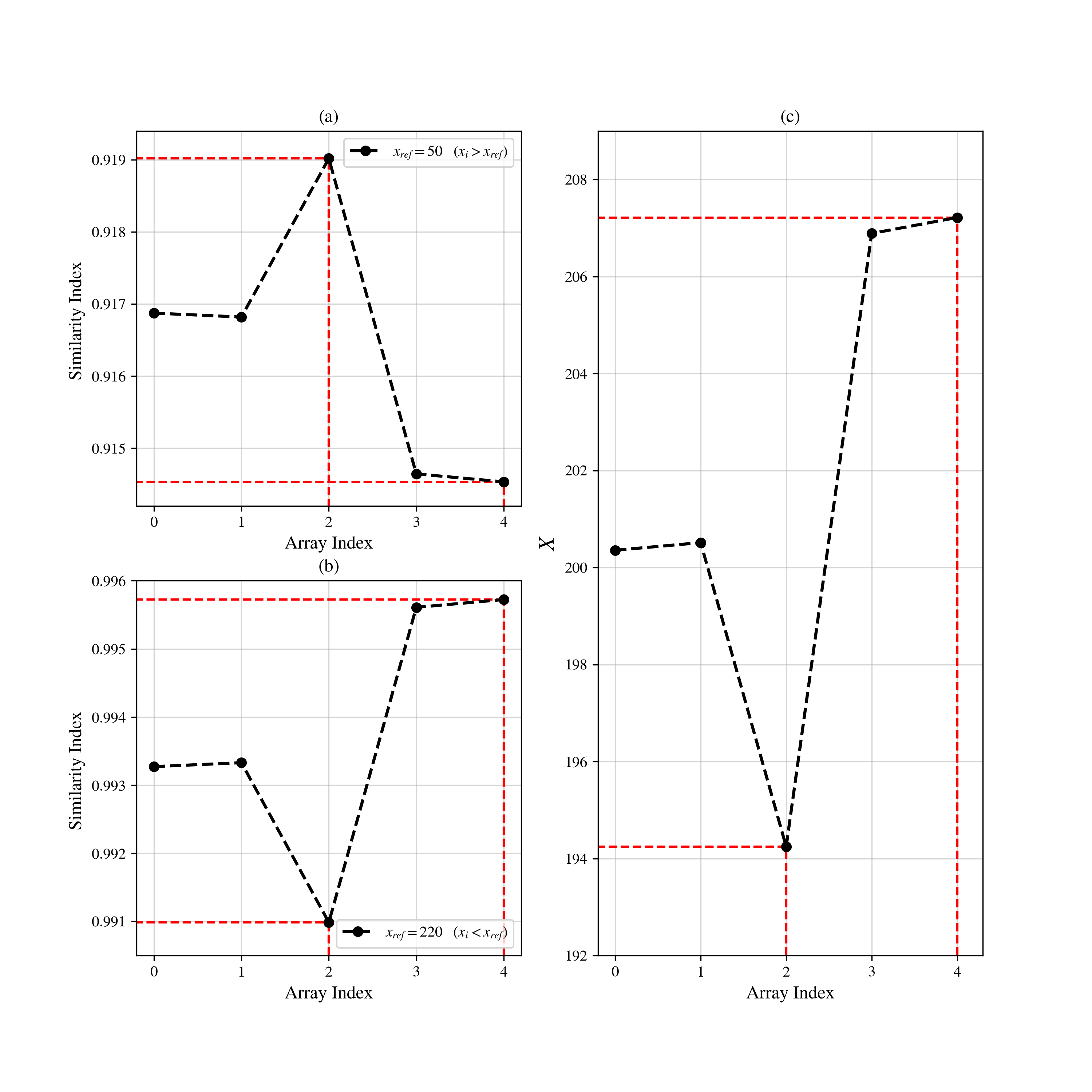} 
\caption{Comparison Plot of Similarity Index. Array Index 2, in the raw data (graph (a)) as well as Similarity Index Satisfying the Inequality \ref{eq:ineq_met} (graph (b)), has the lowest value, while Array Index 4 has the highest value. The opposite is true for graph (a)
}.
\label{fig:1}
\end{figure}

Fig. \ref{fig:1} consists of the raw data, $X$ which is plotted on graph (c), the graph (b) consists of the similarity index satisfying the conditions of inequality in Eq (\ref{eq:ineq_met}) and is in very good agreement with the actual data set (most of the previously published applied data sets (like: Kashyap et al. 2017) did come under this boundary condition). Whereas graph (a) consists of the similarity index dissatisfying the inequality in Eq (\ref{eq:ineq_met}) and satisfying the inequality in Eq(\ref{eq:ineq_not_met}), and hence, is not in proportionality with the raw data, instead, it shows an inverse proportionality with the raw data ($X$). The array index, $2$ and $4$ are highlighted in all the subgraphs for visual verification. 

\begin{figure}[h!]
\centering        
\includegraphics[width=15cm,angle=0]{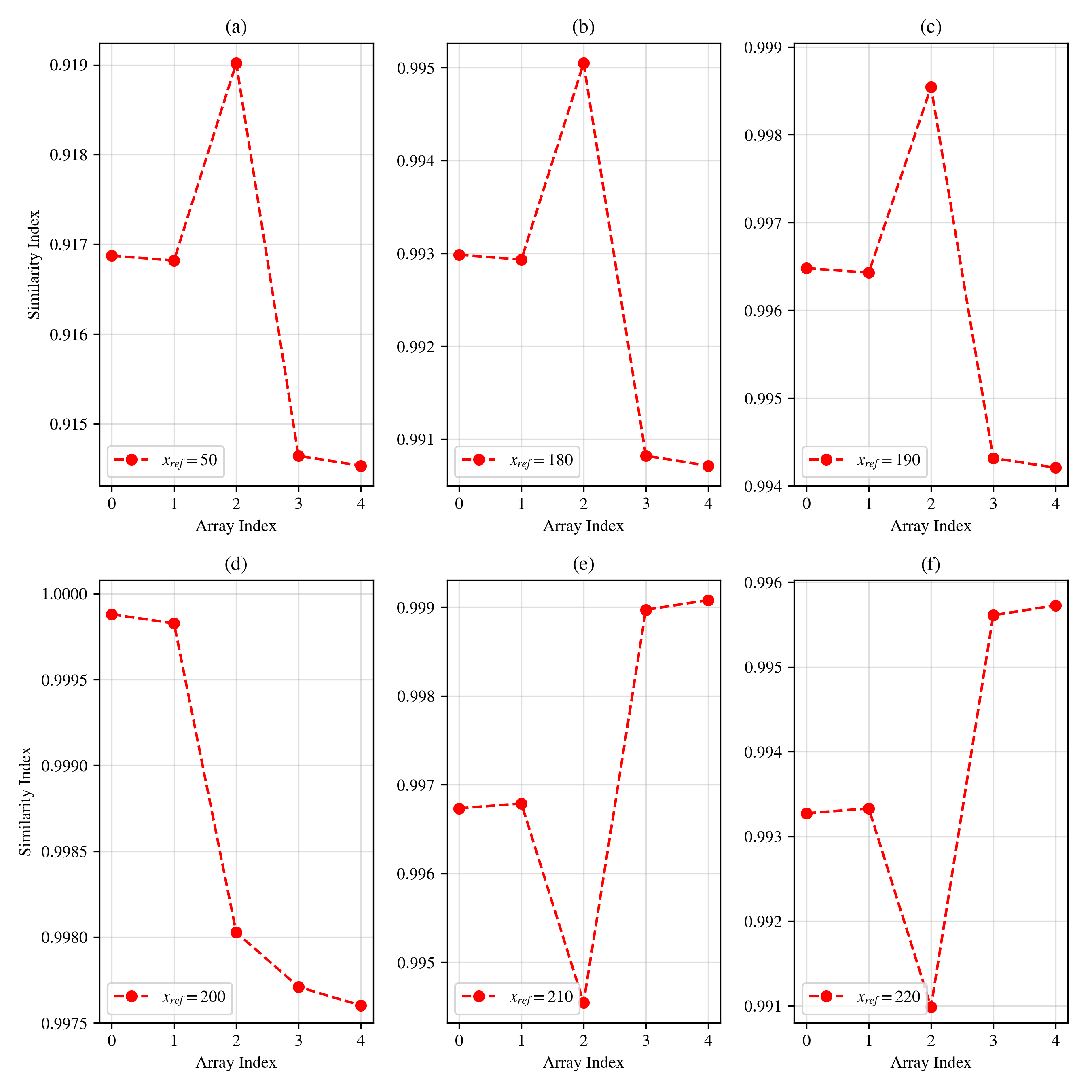} 
\caption{Plot variation in Similarity Index as the value of $x_{ref}$ is ranged from 50 to 220 graph (d) and graph (e) are of special interest here as they encounter an $x_{ref}$ values which have some points in $X$ which are greater than $x_{ref}$ while some points are less than $x_{ref}$.}
\label{fig:3}
\end{figure}

The graphical representation in Figure \ref{fig:1} clearly shows that graph (a) (satisfying the inequality in Eq ({\ref{eq:ineq_not_met}}) is not \textit{directly proportional} with the right hand side actual data. In order to substantiate the claim further, in Fig \ref{fig:3} the Similarity Index is observed as the value of $x_{ref}$ is ranged from $50$ (Eq \ref{eq:ineq_not_met}) to $220$ (Eq \ref{eq:ineq_met}).

The case studies for different ranges of $x_{ref}$ is done in the Figure \ref{fig:3}. Graph (a) is same as graph (a) in Fig \ref{fig:1}. Graph (b), (c) are also similar to graph (a) having $x_{ref}$ value as $180, 190$ respectively, which is still less than all elements of the set $X$. Note that as the value of $x_{ref}$ is increased, the value Similarity Index of respective $x_i \in X$, in comparison is also increased.\\

In graph (d), Array Index 2 corresponds to $194.24$, as shown in Table \ref{tab:primary}, which is clearly less than $x_{ref}$ = 200, and hence the dip is observed. 
Even though $194.24$ is the lowest value in $X$, its corresponding Similarity Index value doesn't yield the least possible value, because except for Array Index 2, all the other Array Indices's value is greater than 200.
In a nutshell, only one element of $X$ satisfies the condition \ref{eq:ineq_met}, which shows direct proportionality with the raw data. The other elements satisfy the condition \ref{eq:ineq_not_met} and hence show inverse proportionality with the set $X$ ($X=S$).

Graph (e) and (f) are similar to graph (c) in Figure \ref{fig:1}, \textit{i.e} all the elements of $X$ satisfy the condition \ref{eq:ineq_met}, all the elements are smaller than $x_{ref}$.

Note that The Graph (e) gives the maximum value for respective similarity index on comparison with Graph (f), despite the fact that the $x_{ref}$ value of Graph (f) is greater than that of Graph (e). Since the dataset $X$ is obtained from the normal distribution, we can say that the optimal Similarity Index value is obtained, when the $x_{ref}$ value is closest to a certain \textit{central} value (in this case,  mean). Graph (b) and (c) also verify the same thing.\\
\newpage
\underline{Verification of the effect of threshold w.r.t similarity index, which has a constant $x_{ref}$ value:}\\

For same dataset, Similarity Index is calculated based on different threshold value using [Eq \ref{SIIeq6}]. Example: Now choose $x_{ref} = 180$, and the corresponding calculations are tabulated in Table 2\\

\begin{center}
	\centering
    \textbf{Table 2: Similarity Index for different threshold values}
    \label{tab:thresh}
	\begin{tabular}{ |l|c|c|c|c|c| } \hline
\diagbox[innerwidth=4cm,innerleftsep=0pt,innerrightsep=0pt]{Threshold}{Array Index}
		& \textbf{0} & \textbf{1} & \textbf{2} & \textbf{3} & \textbf{4}\\\hline
		\textbf{$0.8$} & 0.99298 & 0.99293 & 0.99504 & 0.99082 & 0.99071 \\\hline
        \textbf{$0.7$} & 0.98881 & 0.98872 & 0.99209 & 0.98537 & 0.98520 \\\hline
        \textbf{$0.6$} & 0.98401 & 0.98389 & 0.98869 & 0.97911 & 0.97887 \\\hline
        \textbf{$0.5$} & 0.97837 & 0.97821 & 0.98469 & 0.97176 & 0.97144 \\\hline
        \textbf{$0.4$} & 0.97150 & 0.97129 & 0.97981 & 0.96285 & 0.96242 \\\hline
	\end{tabular}
\end{center}

\begin{figure}[h!]
\centering        
\includegraphics[width=13cm,angle=0]{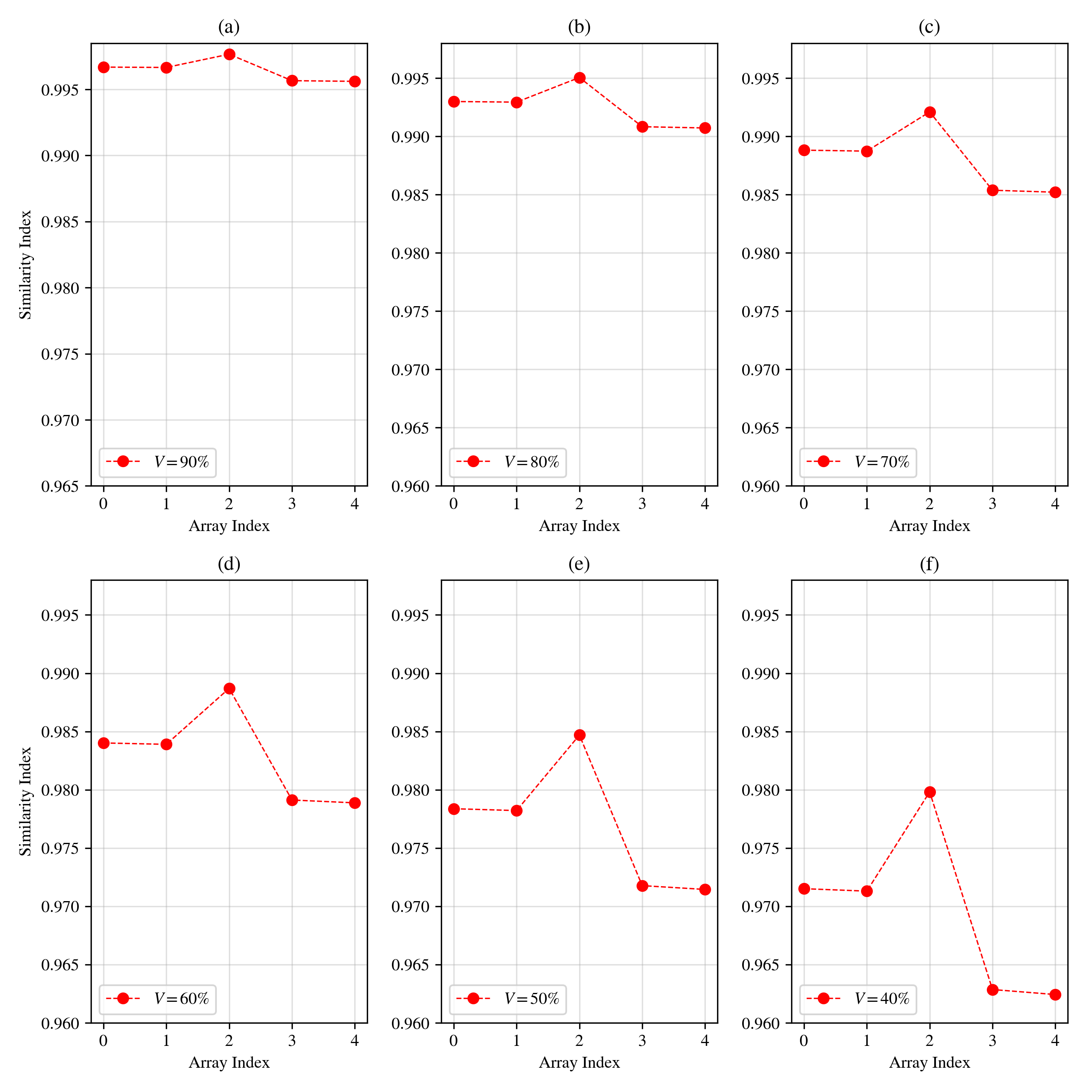} 
\caption{Similarity Index of $X$ with different threshold values. The clustering of points in graph (a), (b), and (c) is because of the limited space in the y-axis (ordinate). The distance among the points in the all the other graphs is the same.}
\label{fig:5}
\end{figure}

Figure \ref{fig:5} is the result of Table 2, and the observations are then made regarding the role threshold value plays with respect to the proportionality of Similarity Index and the set $X=S$. 

As evident, from the graphs in Figure \ref{fig:5}, the threshold value has \textit{no effect} on the proportionality of the Similarity Index with respect to $x_{ref}$. The similarity index value for all the graphs in the figure, irrespective to the threshold value, maintains an inverse proportionality with respect to the $X$. The threshold value only controls the range that will be obtained by the Similarity Index.

\section{Discussion and Conclusion}
We know that a counter example is sufficient to substantiate a fallacy in a previously accepted axiom/postulate. The data set mentioned in this paper are the direct and actual means of examples to see the limitations (or boundary conditions) of similarity index, which is widely used in the application field (Such as: Astrophysics and ecology \citep{schu} and \citep{12}). 
From the above findings, we claim our assumption considered is true, \textit{i.e} the similarity index does not yield the expected result if the conditions in Eq. \ref{eq:ineq_met} is not satisfied or in other words Eq. \ref{eq:ineq_not_met} is satisfied. Figure \ref{fig:1} and Figure \ref{fig:3} verifies the claim made in section 3 through  graphical technique. The effect of threshold value on similarity index was analyzed and verified by Fig \ref{fig:5}, which upholds the claim that threshold value has no impact on the boundary condition for this research findings.
The present and future scope for the application of similarity index in the fields like exo-planets, ecology,... etc, are the motivation to upgrade the similarity index metric tool, with our new boundary conditions. Further research is suggested on why this anomaly is observed? and the mathematical analysis is mandated for the same.

\section{Appendix}
The python source code for this paper is made available on:
 https://github.com/Jar-win/similarity-index-analysis

\section*{Acknowledgments} 
We would like to thank Center Pollution Control Board (CPCB), New Delhi, India, for the atmospheric pollution data. Which intern helped us to find this anomaly.


%
%
\end{document}